\documentclass[12pt,draftclsnofoot,onecolumn]{IEEEtran}

\usepackage{amsmath,amssymb,amsfonts}
\usepackage{graphicx}
\usepackage{lipsum}
\usepackage{color}
\usepackage{tcolorbox}
\usepackage{tabularx}
\usepackage[table]{xcolor}
\usepackage{booktabs}
\usepackage{needspace}
\usepackage{array}

\Needspace{5\baselineskip} 

\bstctlcite{IEEEexample:BSTcontrol}

\newtcolorbox{sidebar}[1][]{
  colback=gray!10,
  colframe=black,
  fonttitle=\bfseries,
  title=#1
}

\newcommand{\mypar}[1]{\vspace{0.2em}\noindent\textbf{#1.} }

\begin{document}

\title{Predictive-State Communication: Innovation Coding and Reconciliation under Delay}

\author{Ozgur~Ercetin, Mohaned~Chraiti \vspace{-1.5\baselineskip}
\thanks{O.~Ercetin  and M.~Chraiti are with the Faculty of Engineering and Natural Sciences, Sabanci University, Istanbul, Turkey (emails: {oercetin,mohaned.chraiti@sabanciuniv.edu}).}%
}

\maketitle

\begin{abstract}
Shannon theory models communication as the reliable transfer of symbol sequences, with performance governed by capacity and rate--distortion limits.  When both endpoints possess strong predictors---as in modern large language models and related generative priors---literal symbol transport is no longer the only operational regime.  We propose \emph{predictive-state communication} (PSC), in which transmitter and receiver maintain an explicit shared predictive state and the physical channel is used primarily to convey \emph{innovations}, i.e., corrective information that reconciles the receiver's provisional trajectory with the transmitter's realized one.  This viewpoint replaces entropy-rate accounting by cross-entropy accounting under model mismatch, and it introduces feasibility constraints that depend jointly on capacity, delay, and perceptual continuity requirements; the resulting operating set is typically a bounded \emph{perception--capacity band} rather than a one-sided threshold.  We outline protocol and architectural implications (state identifiers, anchors, bounded rollback, and patch-based updates), and we give a stylized illustrative example to visualize the induced feasibility region and its dependence on predictive quality.
\end{abstract}

\section{Introduction}
Shannon formalized communication as the reproduction of a selected message at another point, and characterized the maximum reliable rate over a noisy channel by capacity \cite{shannon1948}.
Wyner and collaborators extended this foundation beyond exact reproduction, introducing common information \cite{wyner1975} and rate--distortion with side information \cite{wynerziv1976}.
More recently, semantics-aware communication shifts attention from literal symbol fidelity to the preservation of task-relevant meaning \cite{lu2024semcom}.
Despite their differences, these perspectives share a structural assumption: the receiver is largely \emph{reactive}, reconstructing content only after sufficient information has traversed the link.

This assumption is strained in delay-dominated and interactive settings.
Many signals of interest are highly structured (text, speech, images, sensorimotor streams), yet the space of plausible realizations is enormous.
Under restricted or misspecified symbol models, such streams can appear close to random even when they are predictable under richer latent structure.
If both endpoints can represent that structure and evolve it consistently, then communication can focus on the information required to \emph{select} the intended trajectory among plausible continuations, rather than on transporting every symbol verbatim.

Prediction and compression are tightly linked \cite{witten1987arithmetic}.
In classical source coding this link is captured through conditional probability models; in many modern domains these models are not engineered in closed form but learned.
Large language models (LLMs) exemplify this shift: scaling and training improve predictive fidelity \cite{kaplan2020scaling}, and prediction has already been exploited operationally, e.g., through draft--verify inference schemes such as speculative decoding and speculative sampling \cite{leviathan2023specdec,chen2023specsampling}.
These developments suggest a communication regime in which the channel carries chiefly \emph{innovations}: the discrepancies between what the receiver can predict locally and what the transmitter actually observes or intends.

\emph{Predictive-state communication} (PSC) makes this regime explicit.
In PSC, both transmitter and receiver maintain a predictive state for an evolving process, and the channel is used to convey the innovation required to reconcile these states.
Reliable communication is recast as maintaining state agreement within an application-defined tolerance over time.
The innovation viewpoint is reminiscent of state-space estimation (e.g., Kalman filtering) \cite{welch2001kalman}, but PSC targets high-dimensional learned predictors and, crucially, addresses the operational question that dominates interactive links: what the receiver should do \emph{during delay}.
Rather than gating interaction on symbol arrival, the receiver generates provisional output from its local predictive state and later reconciles it when delayed innovation updates arrive.
In this sense PSC is related to semantic and knowledge-based paradigms that reduce redundancy, but it differs in that it treats \emph{state synchronization under delay} as the primary object (see Table~\ref{tab:comparison}) and it requires explicit reconciliation mechanisms (Fig.~\ref{fig:architecture}).

This paper develops PSC as a communication-theoretic \emph{framework} and protocol viewpoint; it does not claim new coding theorems.
Our contributions are threefold:
(i) we introduce an accounting that characterizes the expected innovation load in terms of cross-entropy (and a mismatch penalty) and use it to motivate a perception--capacity feasibility band;
(ii) we describe architectural primitives for PSC, including state identifiers, anchoring, bounded rollback, and patch-based innovation updates together with mismatch monitoring signals; and
(iii) we present a stylized illustrative example that visualizes the resulting feasibility region and its dependence on predictive quality, delay, and tolerance.

\section{Predictive-State Communication Foundations}
\label{sec:foundations}

\subsection{From symbol transport to innovation transport}
Classical digital communication may be idealized as follows: an encoder observes a source sequence and transmits enough information for the decoder to reproduce that sequence (or an acceptable distortion of it) after a channel delay.
In this view, the receiver is essentially reactive: it waits for the channel output and then reconstructs.

PSC instead treats \emph{prediction during delay} as the default operating mode.
The receiver maintains a local predictive state and generates a \emph{provisional} trajectory while awaiting delayed updates.
The channel is used primarily to convey \emph{innovations}---compact corrective information that reconciles the receiver's provisional trajectory with the transmitter's realized one.
Thus the central quantity is not ``how many symbols per second can be delivered,'' but rather ``how much \emph{unexpected} information per second must be delivered to keep the endpoints aligned within tolerance.''

This does not contradict Shannon theory; it changes the \emph{object being coded}.
Where classical source coding targets the uncertainty in the symbol stream itself, PSC targets the uncertainty \emph{remaining after conditioning on the receiver's predictive state}.
Accordingly, the relevant rate is governed by \emph{cross-entropy} with respect to the receiver's predictor, and model mismatch enters as an explicit penalty term.

\subsection{Objects and notation}
\label{subsec:objects-notation}
We describe PSC at the level of a stochastic process and a shared state abstraction.

\mypar{Committed versus provisional state}
Let $A_t$ denote the most recent \emph{committed} shared anchor at (logical) time $t$; by definition both endpoints agree on $A_t$ once it is committed.
Between anchors, the receiver evolves a \emph{provisional} state $\widetilde S_t$ locally (e.g., by autoregressive generation) while the transmitter evolves its own realized state.
Innovations are transmitted to map the receiver's provisional trajectory back toward the transmitter's realized one, and anchors are used to bound rollback and ensure eventual agreement.

\mypar{Source law and predictor}
Let $X_t$ denote the symbol (token) at time $t$.
Let $H_t$ denote the conditioning information available at the moment $X_t$ is generated \emph{in the committed reference frame}---for example, the last committed anchor $A_t$ together with any agreed protocol metadata.
Write
\[
P(\cdot \mid H_t) \quad \text{for the true conditional law of } X_t \text{ given } H_t,
\]
and write
\[
Q(\cdot \mid H_t) \quad \text{for the receiver's predictive model used for coding/decoding.}
\]
In general $Q$ may be misspecified, and PSC makes this mismatch operational rather than hiding it as ``coding overhead.''

\subsection{Innovation-rate accounting via cross-entropy}
\label{subsec:accounting}
Define the per-step \emph{cross-entropy} (in bits per token)
\begin{equation}
h_t \triangleq H\!\big(P(\cdot\mid H_t),\,Q(\cdot\mid H_t)\big)
= -\sum_x P(x\mid H_t)\log_2 Q(x\mid H_t),
\label{eq:perstep-crossentropy}
\end{equation}
and define the time-average (or stationary) cross-entropy rate
\begin{equation}
\bar h \triangleq \limsup_{n\to\infty}\frac{1}{n}\sum_{t=1}^n h_t.
\label{eq:avg-crossentropy}
\end{equation}
If $(X_t,H_t)$ is stationary and ergodic, then $\bar h$ coincides with the usual cross-entropy rate $H(P,Q)$.

The fundamental decomposition
\begin{equation}
H(P,Q) = H(P) + D_{\mathrm{KL}}(P\|Q)
\label{eq:crossentropy-decompose}
\end{equation}
separates the irreducible uncertainty $H(P)$ from the \emph{mismatch penalty} $D_{\mathrm{KL}}(P\|Q)$.
In PSC this mismatch penalty is not merely a modeling inconvenience: it directly determines the innovation traffic that must traverse the channel to keep the receiver's provisional trajectory aligned.

\mypar{Innovation throughput}
Let $r$ be the generation/consumption rate in tokens per second at the application layer.
Under a standard entropy-coding abstraction (e.g., arithmetic coding) and ignoring protocol overhead for the moment, the required innovation throughput (bits/s) satisfies the baseline estimate
\begin{equation}
R_{\mathrm{innov}} \approx r\,\bar h.
\label{eq:Rinnov}
\end{equation}
To incorporate overhead, let $C$ be the physical link capacity (bits/s), let $\eta\in(0,1)$ be the fraction available for innovation payload after headers, authentication, redundancy, etc., and define
\begin{equation}
C_{\mathrm{innov}} \triangleq \eta C.
\label{eq:Cinnov}
\end{equation}
A necessary feasibility condition is then
\begin{equation}
r\,\bar h \;\le\; C_{\mathrm{innov}}.
\label{eq:capacity-ceiling}
\end{equation}
Equation \eqref{eq:capacity-ceiling} is deliberately an \emph{accounting constraint}, not a coding theorem: it captures the operational fact that better predictors (smaller $\bar h$) reduce required innovation traffic, while mismatch inflates it through \eqref{eq:crossentropy-decompose}.

\subsection{Delay, perceptual continuity, and the perception--capacity band}
\label{subsec:band}
Capacity alone does not characterize PSC viability because PSC is defined by behavior \emph{during delay}.
Let $L$ denote an effective one-way delay in seconds (or an equivalent buffering horizon) over which the receiver must act on provisional output.
Two application-dependent costs become salient:

\begin{itemize}
\item $D_{\mathrm{spec}}(r,L,\ldots)$: a \emph{speculation cost} capturing the harm of acting on provisional output that may later be corrected (e.g., rollback severity, user-perceived inconsistency, or task loss).
\item $D_{\mathrm{starve}}(r,L,\ldots)$: a \emph{starvation cost} capturing the harm of too-slow provisional output (e.g., perceptual stutter, control instability, or missed deadlines).
\end{itemize}
We keep these as explicit design functions because they depend strongly on domain semantics; in Section~\ref{sec:example} we instantiate them in a stylized form purely for visualization.

These costs induce two rate constraints:
\begin{enumerate}
\item an \emph{upper} bound $r \le r_{\max}(L,\text{tolerance})$ imposed by acceptable speculation cost, and
\item a \emph{lower} bound $r \ge r_{\min}(L,\text{tolerance})$ imposed by acceptable starvation cost.
\end{enumerate}
Together with the innovation-capacity ceiling \eqref{eq:capacity-ceiling}, PSC feasibility is naturally a \emph{band} of operating points rather than a single threshold:
\begin{equation}
r_{\min}(L,\cdot)\;\le\; r \;\le\; \min\!\left\{r_{\max}(L,\cdot),\ \frac{C_{\mathrm{innov}}}{\bar h}\right\}.
\label{eq:band}
\end{equation}
When the band is nonempty, PSC can sustain provisional behavior during delay and reconcile later; when it is empty, either the predictor is insufficient (large $\bar h$), the delay is too large for the chosen tolerance, or the physical link is too weak after overhead.

\mypar{Interpretation}
In classical reliable transmission, larger delay is typically a nuisance but not a first-order feasibility parameter.
In PSC, delay directly governs how far provisional trajectories can drift before reconciliation and hence influences both $r_{\max}$ (via speculation) and $r_{\min}$ (via starvation).
Predictive quality enters through $\bar h$: better prediction reduces innovation traffic, enlarging the feasible set through the ceiling term $C_{\mathrm{innov}}/\bar h$.

\subsection{Relation to existing paradigms}
\begin{figure}[t]
    \centering
    \includegraphics[width=0.9\linewidth]{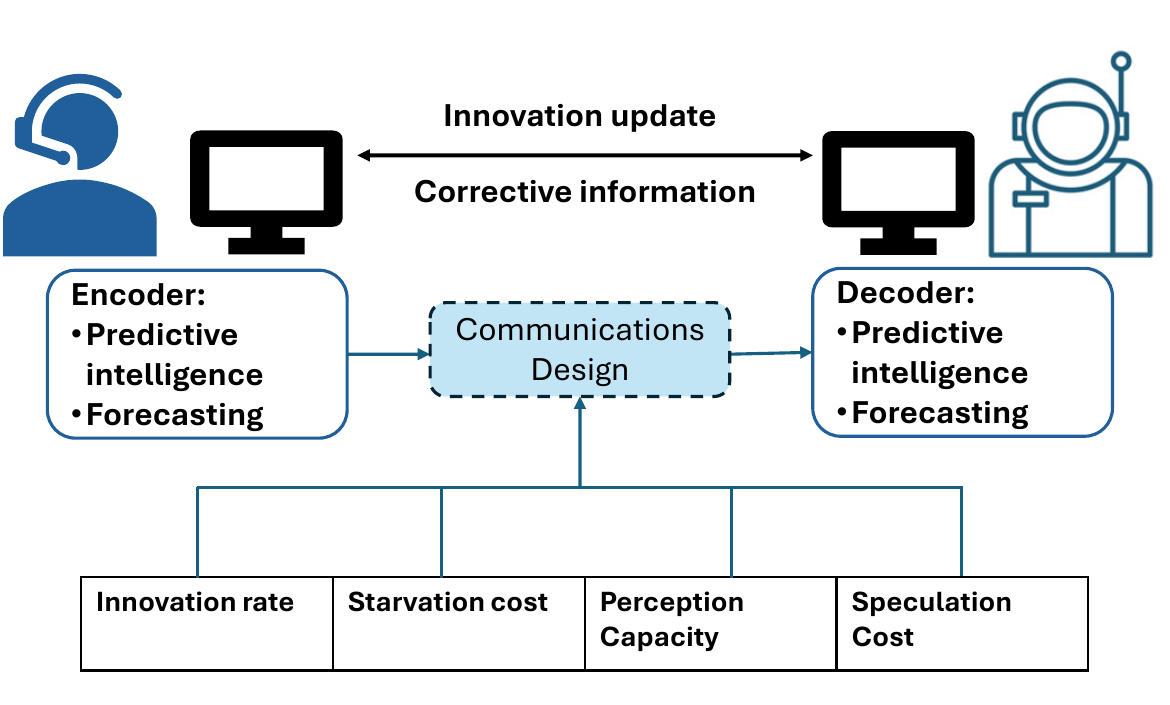}
    \caption{System architecture; the physical channel carries only innovation bits.}
    \label{fig:architecture}
    \vspace{-0.1cm}
\end{figure}

The innovation viewpoint is consistent with classical prediction-based compression, but PSC differs in what it treats as primitive.
PSC elevates three elements to first-class protocol objects:
(i) an explicit shared predictive state (and a way to name it),
(ii) bounded rollback and anchoring for eventual agreement, and
(iii) reconciliation updates whose payload rate is governed by cross-entropy under mismatch.
These distinctions are summarized in Table~\ref{tab:comparison} and operationalized by the architecture in Fig.~\ref{fig:architecture}.

\begin{table*}[t]
\centering
\caption{Comparison of PSC with Related Communication Paradigms}
\label{tab:comparison}
\footnotesize
\begin{tabular}{@{}p{2.5cm}p{2.0cm}p{3.2cm}p{3.25cm}p{4.2cm}@{}}
\toprule
\textbf{Paradigm} & \textbf{Receiver Role} & \textbf{Primary Mechanism} & \textbf{When Receiver Acts} & \textbf{Key Limitation } \\
\midrule
Classical Shannon & Passive reconstructor & Reliable symbol transfer & After symbols arrive & No continuity during delay; user experiences gaps and ignores perceptual quality \\
\midrule
Semantic \cite{lu2024semcom} & Task-bound decoder & Task-relevant feature extraction & After transmission completes & Interaction stalls until compressed features arrive and are decoded \\
\midrule
Knowledge-base \cite{ren2024knowledge} & Generative reconstructor & Shared context combined with generative synthesis & After semantic cues arrive & No explicit mechanism to track drift between endpoint models or resynchronize \\
\midrule
Receiver-centric semantic \cite{tian2024receiver,huang2024receiver} & Requirement specifier & Personalized decoding or feedback loops & During setup; then after each transmission & Round-trip delay gates each request-response exchange \\
\midrule
Generative semantic \cite{yuan2025agi,wang2025diffusion} & Perceptual synthesizer & Diffusion or foundation models for content synthesis & After minimal cues arrive & Synthesis occurs post-transmission; no protocol for state alignment \\
\midrule
Predictive coding and Kalman \cite{welch2001kalman} & State estimator & Innovation transmission for dynamical systems & Continuous (estimation loop) & For low-dimensional explicit models; does not maintain conversational coherence\\
\midrule
\textbf{Predictive-state (this work)} & \textbf{Predictive generator} & \textbf{Continuous prediction combined with innovation patches} & \textbf{During delay (speculative); after patches (reconciliation)} & \textbf{Requires model alignment and bounded speculation windows} \\
\bottomrule
\end{tabular}
\vspace{-0.4cm}
\end{table*}

\section{Architecture and Protocol View}
\label{sec:architecture}

Section~\ref{sec:foundations} cast PSC as innovation transport under an explicit shared predictive state.
We now describe an architectural viewpoint and a minimal set of protocol primitives that realize this abstraction over a real network.
The organizing principle is that PSC is a \emph{state synchronization} problem with two coupled timelines:
(i) a \emph{provisional} timeline, on which the receiver must act immediately using local prediction, and
(ii) a \emph{committed} timeline, on which both endpoints eventually agree via delayed reconciliation.

\subsection{Predictive state and StateID}
\label{subsec:predictive-state}

PSC requires that both endpoints can refer unambiguously to the predictive configuration under which provisional outputs are generated.
To that end we introduce the notion of a predictive state.

\mypar{Definition (Predictive state)}
A \emph{predictive state} at (logical) time $t$ is a tuple
\begin{equation}
S_t \triangleq \big(M,\mathcal{T},\pi,C_t\big),
\label{eq:state-tuple}
\end{equation}
where $M$ is a predictive model (e.g., an LLM or other learned predictor), $\mathcal{T}$ is a deterministic preprocessing/tokenization map, $\pi$ is a generation policy (e.g., decoding rule, temperature, sampling seed conventions), and $C_t$ is the committed context (the agreed history/anchor content up to time $t$).

\mypar{State identifier}
Let $\mathrm{StateID}(S_t)$ denote a collision-resistant identifier for $S_t$ (e.g., a cryptographic digest of versioned descriptors of $M,\mathcal{T},\pi$ together with a digest of $C_t$).
A receiver that uses PSC must be able to verify that incoming reconciliation messages refer to its current committed state:
\begin{equation}
\mathrm{StateID}_{\mathrm{rx}} = \mathrm{StateID}_{\mathrm{tx}}
\quad\text{(committed agreement).}
\label{eq:stateid-agreement}
\end{equation}
When this fails, PSC cannot safely interpret innovations as ``small corrections'' and must fall back to resynchronization (Section~\ref{subsec:resync}).

\mypar{Determinism versus tolerated nondeterminism}
Some predictors are stochastic at inference time.
PSC can be operated in either of two regimes:
(i) \emph{reproducible decoding}, in which all sources of nondeterminism that affect outputs are treated as part of $\pi$ and thus covered by $\mathrm{StateID}$; or
(ii) \emph{nondeterminism-tolerant decoding}, in which the receiver's provisional trajectory is explicitly understood as a hypothesis that will be reconciled by patches.
The protocol must specify which regime is in force, since it affects both patch semantics and the expected innovation load.

\subsection{Anchors, bounded rollback, and commit semantics}
\label{subsec:anchors}

PSC must reconcile provisional behavior with eventual agreement without permitting unbounded rewrites of the past.
This is handled by anchors and bounded rollback.

\mypar{Definition (Anchor)}
An \emph{anchor} is a committed checkpoint of the shared predictive state.
Formally, an anchor at time $t$ is a designated committed context $C_t$ (or its digest) together with $\mathrm{StateID}(S_t)$.
Anchors delimit the region of history subject to correction.

\mypar{Rollback window and speculation horizon}
Fix a rollback window length $W$ tokens (or an equivalent time window).
The receiver is permitted to generate provisional output ahead of the last committed anchor, but only within the window governed by the application tolerance.
Let $H$ denote a speculation horizon (in tokens) that bounds how far ahead the receiver may act without receiving reconciliation.
Operationally, $W$ and $H$ are design parameters that trade user experience (continuity) against the risk and cost of later correction.

\mypar{Commit}
A \emph{commit} operation advances the committed anchor, shrinking the region of history that may be rewritten.
Commits may be periodic, triggered by confidence/mismatch signals, or requested by the application (e.g., at conversational turn boundaries).
Once committed, output prior to the anchor is treated as fixed; corrections that would alter it require an explicit rollback protocol event.

\subsection{Innovation updates as patches}
\label{subsec:patches}

PSC transmits innovations in the form of patches.

\mypar{Definition (Patch)}
A \emph{patch} is a message that reconciles the receiver's provisional trajectory to the transmitter's realized trajectory relative to a specified baseline.
A patch contains:
\begin{enumerate}
\item a baseline reference (the relevant $\mathrm{StateID}$ and anchor digest),
\item a target range in the provisional output to be corrected (e.g., token indices relative to the anchor),
\item an encoded correction payload (which may be a token edit script, a delta on latent variables, or a compressed residual), and
\item optional integrity metadata (authentication tag, checksum, and/or redundancy).
\end{enumerate}

\mypar{Applicability condition}
A patch is \emph{applicable} if and only if the receiver's current committed baseline matches the patch's baseline reference.
If applicability fails, the receiver must not apply the patch, since doing so would corrupt the shared state; instead it initiates resynchronization (Section~\ref{subsec:resync}).

\mypar{Patch semantics}
Applying a patch produces a new committed (or semi-committed) state.
Depending on application tolerance, a patch may:
(i) revise only uncommitted provisional output within the rollback window,
(ii) trigger a rollback event (discarding a portion of provisional output and regenerating), or
(iii) force a hard resync if the correction exceeds tolerance.

\subsection{Mismatch monitoring and control signals}
\label{subsec:mismatch}

PSC performance depends on predictive quality and on the dynamics of mismatch.
It is therefore useful to expose mismatch monitoring signals that inform control of speculation and anchoring.

\mypar{Correction pressure}
Over a window of provisional generation, define correction pressure as a normalized measure of reconciliation load, for example
\begin{equation}
\mathrm{CP} \triangleq \frac{\text{innovation bits delivered in window}}{\text{provisional tokens generated in window}}.
\label{eq:correction-pressure}
\end{equation}
Heuristically, $\mathrm{CP}$ tracks the empirical cross-entropy experienced by the protocol (up to overhead).

\mypar{Rollback pressure}
Define rollback pressure as the frequency (or expected magnitude) with which patches touch deep history within the rollback window.
One simple proxy is the fraction of patches whose edit span intersects the earliest $\rho W$ tokens of the rollback window for some $\rho\in(0,1)$.
High rollback pressure indicates that speculation is running ahead of predictive reliability and suggests shortening $H$, increasing anchor frequency, or switching to a safer mode.

\mypar{Control loop}
PSC naturally supports a control loop:
the receiver adjusts $H$ (speculation horizon) and the transmitter adjusts anchor/commit frequency in response to correction and rollback pressure.
This connects directly to the feasibility band in \eqref{eq:band}: the protocol attempts to choose an operating point $r$ that satisfies the innovation ceiling while maintaining acceptable perceptual continuity.

\subsection{Protocol timeline (informal)}
\label{subsec:timeline}
A typical PSC interaction can be summarized as follows.

\begin{enumerate}
\item \textbf{Initialization.} Endpoints agree on $(M,\mathcal{T},\pi)$ and establish an initial committed context $C_0$ and $\mathrm{StateID}(S_0)$.
\item \textbf{Provisional generation.} The receiver generates provisional tokens/events using its local predictive state while awaiting delayed updates.
\item \textbf{Innovation transmission.} The transmitter sends patches referencing the current baseline and anchor, carrying the innovation needed to reconcile.
\item \textbf{Reconciliation.} The receiver applies applicable patches, possibly rolling back and regenerating within the allowed window.
\item \textbf{Commit/anchor update.} Once provisional divergence is controlled, the transmitter issues a commit advancing the anchor, thereby fixing a new baseline.
\item \textbf{Adaptation.} Mismatch signals adjust horizon and anchor frequency; extreme mismatch triggers resynchronization.
\end{enumerate}

This timeline is intentionally abstract: PSC specifies the \emph{objects} (states, anchors, patches) and correctness conditions (applicability, bounded rollback), while leaving the representation of corrections and the application tolerance model to the system designer.

\subsection{Resynchronization and safety considerations}
\label{subsec:resync}
PSC must handle failure modes in which the shared baseline is lost or mismatch becomes too large.

\mypar{Resynchronization}
Resync is invoked when:
(i) StateID disagreement is detected, (ii) patch applicability fails repeatedly, or (iii) rollback pressure exceeds tolerance.
A resync protocol may transmit a fresh anchor (or a compact summary sufficient to reconstruct $C_t$), re-establish $\mathrm{StateID}$ agreement, and restart provisional generation.

\mypar{Integrity and provenance}
Because PSC alters the receiver's provisional trajectory after the fact, integrity checks are essential.
At minimum, patch messages should be authenticated, and anchors should be committed with verifiable digests.
For applications with audit requirements, the system may log the sequence of anchors and patches so that committed output can be reconstructed and verified.

\mypar{When PSC is appropriate}
PSC is most compelling when delay is large relative to interaction timescales and when predictors are strong enough that $\bar h$ is small (cf.\ \eqref{eq:Rinnov}--\eqref{eq:band}).
When predictors are weak or mismatch is persistent, PSC may degrade to frequent rollbacks or resync, in which case conventional transmission or hybrid schemes may be preferable.

\begin{figure*}[!ht]
    \centering
    \includegraphics[width=0.9\textwidth,height=5cm]{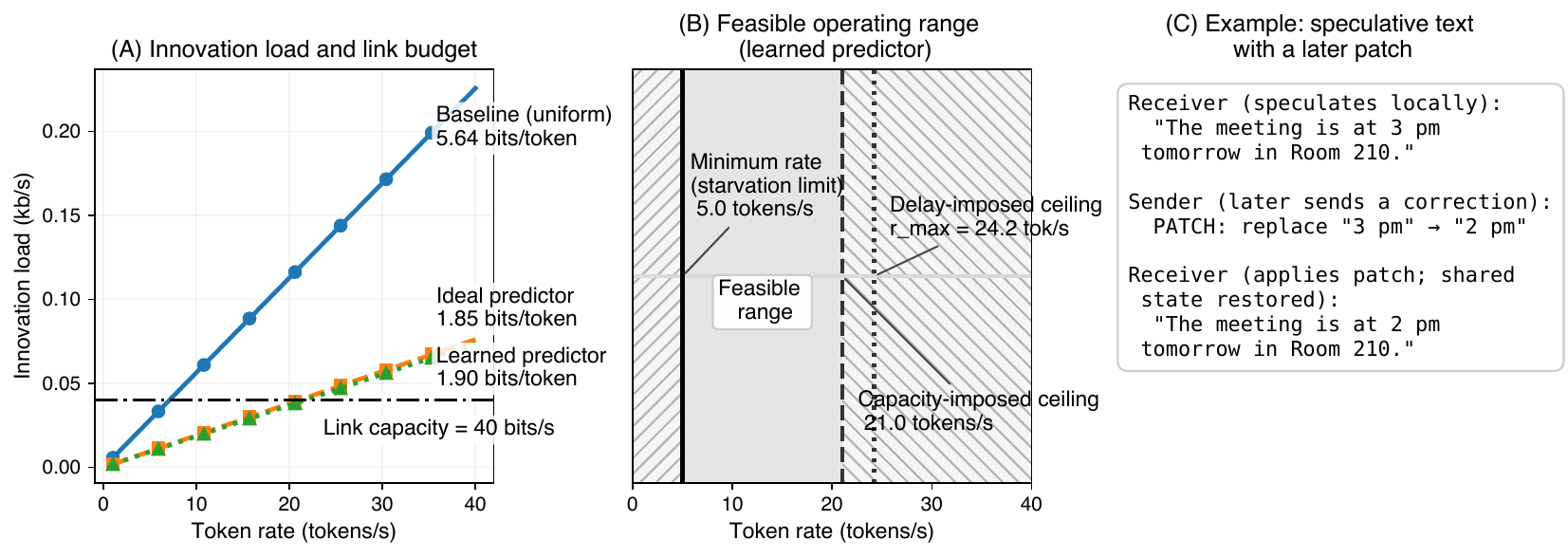}
    \caption{Illustration of PSC.
(A) Innovation load versus token rate for three predictors against a fixed link budget; better prediction lowers innovation load.
(B) Operating interval for the learned predictor between starvation floor and capacity/delay ceiling; shaded region shows feasible rates.
(C) Operation schematic: receiver speculates for continuity; sender transmits compact patch to restore agreement without full retransmission.}
    \label{fig:exp}
\end{figure*}

\section{Illustrative Example: Visualizing the Feasibility Band}
\label{sec:example}

This section is \emph{illustrative, not evaluative}.
Our aim is to visualize how predictive quality (through $\bar h$), link capacity, and delay-sensitive tolerances interact through \eqref{eq:band}.
To keep the exposition transparent, we use simple proxy quantities and a standard cross-entropy computation from language modeling.

\subsection{Setup: cross-entropy as innovation-rate proxy}
\label{subsec:example-setup}

We consider token streams and measure predictive quality by cross-entropy in bits per token.
Let $Q$ be a fixed pretrained language model used as the receiver's predictor; for a corpus with token sequence $x_{1:n}$, define the empirical cross-entropy
\begin{equation}
\widehat h \triangleq -\frac{1}{n}\sum_{t=1}^n \log_2 Q(x_t \mid x_{<t}),
\label{eq:empirical-crossentropy}
\end{equation}
computed in the usual teacher-forced manner.
Under the accounting view of Section~\ref{subsec:accounting}, a baseline innovation throughput estimate is then
\begin{equation}
\widehat R_{\mathrm{innov}} \approx r\,\widehat h.
\label{eq:empirical-Rinnov}
\end{equation}
We emphasize that \eqref{eq:empirical-Rinnov} is a proxy: it ignores protocol overhead, ignores patch structure, and treats the predictor's token probabilities as directly usable for entropy coding.

\subsection{Toy delay-tolerance model for visualization}
\label{subsec:toy-costs}

To visualize the effect of delay, we require explicit forms for the speculation and starvation costs introduced in Section~\ref{subsec:band}.
We adopt the following \emph{toy} instantiation, used only to produce the plots in Fig.~\ref{fig:exp}:
\begin{align}
D_{\mathrm{starve}}(r; r_0) &\triangleq \mathbf{1}\{r < r_0\},
\label{eq:toy-starve}\\
D_{\mathrm{spec}}(r,L; \kappa) &\triangleq \mathbf{1}\{rL > \kappa\}.
\label{eq:toy-spec}
\end{align}
Here $r_0$ is a minimum continuity rate (tokens/s) and $\kappa$ is a maximum tolerated \emph{speculation depth} in tokens.
Thus $D_{\mathrm{starve}}=1$ indicates unacceptable stutter, and $D_{\mathrm{spec}}=1$ indicates that delay would permit too many provisional tokens to accumulate before reconciliation, exceeding tolerance.
Under \eqref{eq:toy-starve}--\eqref{eq:toy-spec}, the feasibility band \eqref{eq:band} reduces to
\begin{equation}
r_0 \;\le\; r \;\le\; \min\!\left\{\frac{\kappa}{L},\ \frac{C_{\mathrm{innov}}}{\bar h}\right\}.
\label{eq:toy-band}
\end{equation}
More refined models could make $D_{\mathrm{spec}}$ and $D_{\mathrm{starve}}$ continuous and task-specific; the qualitative role of delay and predictive quality remains the same.

\subsection{Accounting example with a stylized Markov source}
\label{subsec:markov-toy}

We first illustrate the distinction between entropy and cross-entropy using a stylized bigram (order-1 Markov) source.
Let $P$ be the true bigram model and let $Q$ be a mismatched predictor.
The entropy rate $H(P)$ quantifies the irreducible uncertainty of the source, while $H(P,Q)$ governs the expected code length when coding according to $Q$.
By \eqref{eq:crossentropy-decompose}, the penalty $D_{\mathrm{KL}}(P\|Q)$ appears additively and can be interpreted as the additional innovation traffic incurred by mismatch.
This toy example supports the conceptual point: improving predictive quality reduces the innovation ceiling term $C_{\mathrm{innov}}/\bar h$ in \eqref{eq:band}, thereby enlarging the feasible region.

\subsection{Language-model cross-entropy and feasibility band visualization}
\label{subsec:wikitext}

We next compute empirical cross-entropies on WikiText-2 to obtain plausible values of $\widehat h$ for modern predictors.
We tokenize the corpus using a GPT-2 BPE tokenizer and evaluate the negative log-likelihood in a standard sliding-window fashion, yielding $\widehat h$ in bits/token.
Table~\ref{tab:results} reports representative values for different predictors and uses \eqref{eq:empirical-Rinnov} to translate them into innovation throughput at a nominal generation rate.

To visualize the band, fix a link innovation capacity $C_{\mathrm{innov}}$ and delay $L$, and choose toy tolerance parameters $(r_0,\kappa)$.
Equation~\eqref{eq:toy-band} then defines an interval of permissible rates.
Fig.~\ref{fig:exp} depicts this interval as a region in the $(\bar h,r)$ plane: improving prediction (smaller $\bar h$) increases the innovation ceiling $C_{\mathrm{innov}}/\bar h$ and can turn an empty band into a nonempty one.

\mypar{Interpretation and limitations}
The numbers in Table~\ref{tab:results} and the regions in Fig.~\ref{fig:exp} should be read as \emph{accounting-level} illustrations.
A deployed PSC system must account for protocol overhead, patch representation efficiency, integrity checks, and the fact that corrections may arrive in bursts rather than at a steady rate.
Nevertheless, the example highlights the central PSC claim: predictive intelligence changes which links are operationally viable for interactive use, because the channel need only carry innovations rather than full symbol streams.

\begin{table}[t]
\centering
\caption{Resulting cross-entropy and feasible token-rate ranges on WikiText-2 under fixed channel and delay.}
\begin{tabular}{lccc}
\toprule
\textbf{Predictor} & Cross-entropy & Capacity-limited & Feasible Range \\
& [bits/token] & rate [tokens/s] & [tokens/s] \\
\midrule
GPT-2      & 4.68  & 8.5 & [5.0, 8.5] \\
Unigram    & 10.40 & 3.8 & [5.0, 3.8] \\
Uniform    & 15.62 & 2.6 & [5.0, 2.6] \\
\bottomrule
\end{tabular}
\label{tab:results}
\end{table}
\vspace{-0.2cm}

\section{Research Challenges and Open Directions}
\label{sec:challenges}

PSC shifts the central engineering question from ``how do we transport symbols reliably?'' to ``how do we maintain state agreement within tolerance while acting during delay?''
This section enumerates research directions that arise when prediction, coding, and protocol design are coupled through the feasibility band \eqref{eq:band}.

\subsection{State specification and interoperability}
\label{subsec:interoperability}

PSC depends on endpoints sharing a predictive state definition precise enough to make innovations meaningful.
This raises questions of standardization:
what exactly is covered by $\mathrm{StateID}$ (model versioning, tokenizer details, decoding policy, randomness conventions),
how contexts are serialized and digested,
and how backward compatibility is handled as models evolve.
A practical PSC ecosystem likely requires negotiated profiles (analogous to codec profiles), where different StateID regimes trade reproducibility against flexibility.

\subsection{Patch representations and optimal innovation coding}
\label{subsec:patch-representations}

The accounting view in Section~\ref{subsec:accounting} treats innovations as if they were coded at cross-entropy rate.
A real protocol must represent corrections in some form:
token edit scripts, latent-variable deltas, semantic constraints, or hybrid payloads.
A basic open problem is to characterize which representations approach the cross-entropy limit under realistic constraints (bounded rollback, integrity metadata, burstiness) and how close practical patch codecs can get.
Related questions include:
how to compress patches jointly across time,
how to exploit structure in mismatch events,
and how to schedule patch transmission under capacity fluctuations.

\subsection{Control under delay: adapting speculation and anchoring}
\label{subsec:control}

PSC operation requires choosing a speculation horizon $H$, a rollback window $W$, and an anchor/commit policy.
These choices naturally form a control problem driven by mismatch signals such as correction pressure \eqref{eq:correction-pressure} and rollback pressure.
Key questions include:
how to detect regime shifts (sudden mismatch increases),
how to trade short-term continuity against long-term correction cost,
and how to guarantee stability (avoid oscillations between over-speculation and excessive anchoring).
Integrating this control loop with congestion control and transport protocols is an additional challenge.

\subsection{Security, provenance, and adversarial behavior}
\label{subsec:security}

Because PSC permits post hoc modification of provisional output, integrity and provenance are central.
Patch authentication and anchor digests are minimum requirements, but stronger mechanisms may be needed in adversarial environments:
replay protection, denial-of-service resilience, and safeguards against malicious patches that induce harmful rollbacks or subtly corrupt the shared state.
For safety-critical applications, it may be necessary to log anchors and patches to support auditability and non-repudiation.

\subsection{Evaluation methodology}
\label{subsec:evaluation}

PSC is not well evaluated by conventional link-layer metrics alone.
Beyond throughput and error rate, a PSC evaluation should report:
(i) innovation payload rate and overhead,
(ii) correction and rollback pressure,
(iii) user- or task-level continuity measures tied to $D_{\mathrm{starve}}$,
and (iv) correction severity measures tied to $D_{\mathrm{spec}}$.
Constructing benchmarks that expose delay sensitivity and that stress mismatch (domain shifts, context truncation, model updates) is an open methodological need.

\subsection{When PSC helps and when it does not}
\label{subsec:when-helps}

PSC is most attractive when predictors are strong (small $\bar h$), delay is consequential, and the application can tolerate bounded post hoc correction.
When predictors are weak or mismatch is persistent, PSC can devolve into frequent rollbacks or resynchronization, reducing to conventional transport with extra overhead.
Characterizing these regimes---and designing graceful fallbacks and hybrids---is important for responsible deployment.

\section{Conclusion}
PSC reframes communication for interactive, delay-dominated settings by making prediction and reconciliation explicit protocol objects.
In this framework, innovation traffic is governed by cross-entropy under mismatch, and feasibility is determined by a bounded perception--capacity band that depends jointly on predictive quality, capacity, and delay-sensitive tolerances.
We outlined minimal protocol primitives (StateID, anchors, bounded rollback, patches) and used a stylized example to visualize the induced feasibility region.
Future work includes patch coding, control of speculation and anchoring, interoperability standards for predictive state, and evaluation methodologies that capture continuity and correction dynamics.

\bibliographystyle{IEEEtran}
\bibliography{IEEEabrv,ref3}

\end{document}